\documentclass{hep99}
\begin{document}


\begin{titlepage}
 
\begin{flushright}
CERN-TH/99-243\\
hep-ph/9908341
\end{flushright}
 
\vspace{2.5cm}
 
\begin{center}
\Large\bf New Strategies to Extract CKM Phases from\\ 
\vspace*{0.3truecm}
Non-Leptonic \boldmath $B$ Decays\unboldmath
\end{center}

\vspace{1.8cm}
 
\begin{center}
Robert Fleischer\\
{\sl Theory Division, CERN, CH-1211 Geneva 23, Switzerland}
\end{center}
 
\vspace{2.3cm}
 
\begin{center}
{\bf Abstract}\\[0.3cm]
\parbox{11cm}{Several new strategies to extract CKM phases, including
a determination of $\gamma$ from $B_{s(d)}\to J/\psi\, K_{\rm S}$ decays 
and a general approach, which makes use of the angular distributions of 
certain $B_{d,s}$ modes, such as $B_d\to J/\psi\,\rho^0$ and $B_s\to 
J/\psi\,\phi$, are discussed. Special emphasis is put on a simultaneous 
determination of $\beta$ and $\gamma$ with the help of the decays 
$B_d\to \pi^+\pi^-$ and $B_s\to K^+K^-$, which relies only on the $U$-spin 
flavour symmetry of strong interactions and is not affected by any penguin 
or final-state-interaction effects.
}
\end{center}
 
\vspace{2.2cm}
 
\begin{center}
{\sl Talk given at the\\
International Europhysics Conference on High Energy Physics -- 
EPS-HEP '99,\\
Tampere, Finland, 15--21 July 1999\\ 
To appear in the Proceedings}
\end{center}
 
\vspace{1.5cm}
 
\vfil
\noindent
CERN-TH/99-243\\
August 1999
 
\end{titlepage}
 
\thispagestyle{empty}
\vbox{}
\newpage
 
\setcounter{page}{0}
 

\title{New strategies to extract CKM phases from non-leptonic B decays}

\author{Robert Fleischer}
%

\address{CERN, Theory Division, CH-1211 Geneva 23, Switzerland\\[3pt]
E-mail: {\tt Robert.Fleischer@cern.ch}}

\abstract{Several new strategies to extract CKM phases, including
a determination of $\gamma$ from $B_{s(d)}\to J/\psi\, K_{\rm S}$ decays 
and a general approach, which makes use of the angular distributions of 
certain $B_{d,s}$ modes, such as $B_d\to J/\psi\,\rho^0$ and $B_s\to 
J/\psi\,\phi$, are discussed. Special emphasis is put on a simultaneous 
determination of $\beta$ and $\gamma$ with the help of the decays 
$B_d\to \pi^+\pi^-$ and $B_s\to K^+K^-$, which relies only on the $U$-spin 
flavour symmetry of strong interactions and is not affected by any penguin 
or final-state-interaction effects.} 

\maketitle

\section{Introduction}

Among the central targets of future $B$-physics experiments is the direct 
measurement of the three angles $\alpha$, $\beta$ and $\gamma$ of the 
unitarity triangle of the Cabibbo--Kobayashi--Maskawa (CKM) matrix.
However, only the extraction of $\beta$ with the help of the ``gold-plated'' 
mode $B_d\to J/\psi K_{\rm S}$ is quite straightforward. In the test of the 
Standard-Model description of CP violation, the determination of the 
angle $\gamma$ is a crucial element. Since the $e^+e^-$ $B$-factories 
operating at the $\Upsilon(4S)$ resonance will not be in a position to 
explore $B_s$ decays, a strong emphasis was given so far, in the literature, 
to decays of non-strange $B$-mesons. However, also the $B_s$ system provides 
interesting strategies to determine $\gamma$, which appear promising for 
dedicated $B$-physics experiments at hadron machines, such as LHCb (CERN) 
or BTeV (Fermilab). The new strategies, discussed here, are---in contrast 
to clean strategies using pure ``tree'' decays, such as 
$B_s\to D_s^\pm K^\mp$---very sensitive to new-physics contributions to 
the corresponding decay amplitudes and may play an important role to
explore the physics beyond the Standard Model.

\section{Extracting $\gamma$ from $B_{s(d)}\to J/\psi\, K_{\rm S}$}

The decays $B_s\to J/\psi\, K_{\rm S}$ and $B_d\to J/\psi\, K_{\rm S}$
are related to each other by interchanging all down and strange quarks,
i.e.\ through the $U$-spin flavour symmetry of strong interactions. 
Whereas the CP-violating weak phase factor $e^{i\gamma}$ is strongly 
Cabibbo-suppressed in the decay amplitude of the ``gold-plated'' mode 
$B_d\to J/\psi\, K_{\rm S}$, this is not the case in $B_s\to J/\psi\, 
K_{\rm S}$. Consequently, there may be sizeable CP-violating effects 
in this channel, which are due to certain penguin topologies. If we make 
use of the $U$-spin flavour symmetry, the CKM angle $\gamma$ and 
interesting hadronic quantities can be extracted by combining the 
``direct'' and ``mixing-induced'' CP asymmetries 
${\cal A}_{\rm CP}^{\rm dir}(B_s\to J/\psi\, K_{\rm S})$ and 
${\cal A}_{\rm CP}^{\rm mix}(B_s\to J/\psi\, K_{\rm S})$ with the 
CP-averaged $B_{d(s)}\to J/\psi\, K_{\rm S}$ branching ratios \cite{BsPsiK}.
Remarkably, the theoretical accuracy of this approach is only limited 
by $U$-spin-breaking corrections. In particular, there are no problems 
due to final-state-interaction (FSI) effects. An interesting by-product
of this  strategy is that it allows us to take into account the---presumably 
very small---penguin contributions in the determination of the
$B^0_d$--$\overline{B^0_d}$ mixing phase $\phi_d=2\beta$ from 
$B_d\to J/\psi\, K_{\rm S}$, which is an important issue in view of the 
impressive accuracy that can be achieved with second-generation 
$B$-physics experiments. Moreover, we have an interesting relation between 
the direct $B_{s(d)}\to J/\psi\, K_{\rm S}$ CP asymmetries and the 
corresponding CP-averaged branching ratios:
\begin{equation}
\frac{{\cal A}_{\rm CP}^{\rm dir}(B_d\to J/\psi\, 
K_{\rm S})}{{\cal A}_{\rm CP}^{\rm dir}(B_s\to J/\psi\, K_{\rm S})}\approx
-\,\frac{\mbox{BR}(B_s\to J/\psi\, K_{\rm S})}{\mbox{BR}(B_d\to J/\psi\, 
K_{\rm S})}\,.
\end{equation}
The experimental feasibility of the extraction of $\gamma$ sketched above
depends strongly on the size of the penguin effects in 
$B_s\to J/\psi\, K_{\rm S}$, which are very hard to estimate. A similar
strategy is provided by $B_{d (s)}\to D^{\,+}_{d(s)}\, D^{\,-}_{d(s)}$ 
decays. For a detailed discussion, the reader is referred to \cite{BsPsiK}.

\section{Extracting $\beta$ and $\gamma$ from the decays 
$B_d\to \pi^+\pi^-$ and $B_s\to K^+K^-$}

In the literature, $B_d\to\pi^+\pi^-$ usually appears as a tool to 
probe $\alpha=180^\circ-\beta-\gamma$. However, penguin contributions 
preclude a reliable determination of $\alpha$ from the CP-violating
observables of the decay $B_d\to\pi^+\pi^-$ that arise in the usual 
time-dependent CP asymmetry
\begin{equation}
a_{\rm CP}(t)={\cal A}_{\rm CP}^{\rm dir}\cos(\Delta M_d t)+
{\cal A}_{\rm CP}^{\rm mix}\sin(\Delta M_d t).
\end{equation}
Although several strategies were proposed to control these penguin 
uncertainties, they are usually very challenging from an experimental 
point of view.

In the following, a new way of using the CP-violating observables 
of $B_d\to\pi^+\pi^-$ is discussed \cite{BsKK}: combining them 
with those of $B_s\to K^+K^-$---the $U$-spin counterpart of 
$B_d\to\pi^+\pi^-$---a simultaneous determination of $\phi_d=2\beta$ and 
$\gamma$ becomes possible. This approach is not affected by any penguin 
topologies---it rather makes use of them---and does not rely on 
certain ``plausible'' dynamical or model-dependent assumptions. Moreover, 
FSI effects, which attracted considerable attention in the recent literature 
in the context of the determination of $\gamma$ from $B\to\pi K$ decays, 
do not lead to any problems, and the theoretical accuracy is only limited 
by $U$-spin-breaking effects. This strategy, which is also very promising 
to search for indications of new physics \cite{FMat}, is conceptually quite 
similar to the extraction of $\gamma$ from $B_{s(d)}\to J/\psi\, K_{\rm S}$ 
discussed in the previous subsection. However, it appears to be more 
favourable in view of the $U$-spin-breaking effects and the experimental 
feasibility.

If we make use of the unitarity of the CKM matrix and apply the Wolfenstein 
parametrization, generalized to include non-leading terms in 
$\lambda\equiv|V_{us}|=0.22$, the $B_d^0\to\pi^+\pi^-$ 
decay amplitude can be expressed as follows \cite{BsKK}:
\begin{equation}\label{Bdpipi-ampl}
A(B_d^0\to\pi^+\pi^-)=e^{i\gamma}\left(1-\frac{\lambda^2}{2}\right){\cal C}
\left[1-d\,e^{i\theta}e^{-i\gamma}\right],
\end{equation}
where
\begin{equation}\label{C-def}
{\cal C}\equiv\lambda^3A\,R_b\left(A_{\rm cc}^{u}+A_{\rm pen}^{ut}\right),
\end{equation}
with $A_{\rm pen}^{ut}\equiv A_{\rm pen}^{u}-A_{\rm pen}^{t}$, and
\begin{equation}\label{d-def}
d\,e^{i\theta}\equiv\frac{1}{(1-\lambda^2/2)R_b}
\left(\frac{A_{\rm pen}^{ct}}{A_{\rm cc}^{u}+A_{\rm pen}^{ut}}\right).
\end{equation}
Here $A_{\rm cc}^{u}$ is due to current--current contributions, whereas the 
amplitudes $A_{\rm pen}^{j}$ describe penguin topologies with internal $j$ 
quarks ($j\in\{u,c,t\})$. The relevant CKM factors are given by 
$A\equiv|V_{cb}|/\lambda^2$ and $R_b\equiv|V_{ub}/(\lambda
V_{cb})|$. In analogy to (\ref{Bdpipi-ampl}), the $B_s^0\to K^+K^-$
decay amplitude can be parametrized as
\begin{equation}\label{BsKK-ampl}
A(B_s^0\to K^+K^-)=e^{i\gamma}\lambda\,{\cal C}'\left[1+\frac{1}{\varepsilon}
\,d'e^{i\theta'}e^{-i\gamma}\right],
\end{equation}
where 
\begin{equation}
{\cal C}'\equiv\lambda^3A\,R_b\left(A_{\rm cc}^{u'}+A_{\rm pen}^{ut'}\right)
\end{equation}
and 
\begin{equation}\label{dp-def}
d'e^{i\theta'}\equiv\frac{1}{(1-\lambda^2/2)R_b}
\left(\frac{A_{\rm pen}^{ct'}}{A_{\rm cc}^{u'}+A_{\rm pen}^{ut'}}\right)
\end{equation}
correspond to (\ref{C-def}) and (\ref{d-def}), respectively, and 
$\varepsilon\equiv\lambda^2/(1-\lambda^2)$. 

The decays $B_d\to\pi^+\pi^-$ and $B_s\to K^+K^-$ are related to each 
other by interchanging all down and strange quarks. Consequently, the 
$U$-spin flavour symmetry of strong interactions implies
\begin{equation}\label{U-spin-rel}
d'=d\quad\mbox{and}\quad\theta'=\theta.
\end{equation}
If we assume that the $B^0_s$--$\overline{B^0_s}$ mixing phase $\phi_s$ 
is negligibly small, as expected in the Standard Model, or that it is 
fixed through $B_s\to J/\psi\,\phi$ (see, for example, \cite{ddf1}), the 
four CP-violating observables provided by $B_d\to\pi^+\pi^-$ and 
$B_s\to K^+K^-$ depend---in the strict $U$-spin limit---on the 
four ``unknowns'' $d$, $\theta$, $\phi_d=2\beta$ and $\gamma$. We 
therefore have sufficient observables at our disposal to extract these 
quantities simultaneously. In order to determine $\gamma$, it suffices 
to consider ${\cal A}_{\rm CP}^{\rm mix}(B_s\to K^+K^-)$ and the direct 
CP asymmetries ${\cal A}_{\rm CP}^{\rm dir}(B_s\to K^+K^-)$, 
${\cal A}_{\rm CP}^{\rm dir}(B_d\to\pi^+\pi^-)$. If we make use, in addition,
of ${\cal A}_{\rm CP}^{\rm mix}(B_d\to\pi^+\pi^-)$, $\phi_d$ can be determined 
as well. The formulae to implement this approach in a mathematical way
are given in \cite{BsKK}.

If we use the $B^0_d$--$\overline{B^0_d}$ mixing phase as an input, 
there is a different way of combining ${\cal A}_{\rm CP}^{\rm dir}
(B_d\to\pi^+\pi^-)$, ${\cal A}_{\rm CP}^{\rm mix}(B_d\to\pi^+\pi^-)$ with 
${\cal A}_{\rm CP}^{\rm dir}(B_s\to K^+K^-)$, 
${\cal A}_{\rm CP}^{\rm mix}(B_s\to K^+K^-)$. The point is that these
observables allow us to fix contours in the $\gamma$--$d$ and 
$\gamma$--$d'$ planes as functions of the $B^0_d$--$\overline{B^0_d}$ and 
$B^0_s$--$\overline{B^0_s}$ mixing phases in a {\it theoretically clean} 
way. In order to extract $\gamma$ and the hadronic parameters $d$, $\theta$, 
$\theta'$ with the help of these contours, the $U$-spin relation $d'=d$ is
sufficient. An illustration of this approach for a specific example can be 
found in \cite{BsKK}. A first experimental feasibility study for LHCb, 
using the same set of observables, gave an uncertainty of 
$\left.\Delta\gamma\right|_{\rm exp}=2.3^\circ$ for five years of 
data taking and looks very promising \cite{wilkinson}.

It should be emphasized that the theoretical accuracy of $\gamma$ and
of the hadronic parameters $d$, $\theta$ and $\theta'$ is only limited
by $U$-spin-breaking effects. In particular, it is not affected by
any FSI or penguin effects. A first consistency check is provided by 
$\theta=\theta'$. Moreover, we may determine the normalization
factors ${\cal C}$ and ${\cal C}'$ of the $B^0_d\to\pi^+\pi^-$ and
$B^0_s\to K^+K^-$ decay amplitudes (see (\ref{Bdpipi-ampl}) and 
(\ref{BsKK-ampl})) with the help of the corresponding CP-averaged
branching ratios. Comparing them with the ``factorized'' result
\begin{equation}
\left|\frac{{\cal C}'}{{\cal C}}\right|_{\rm fact}=\,
\frac{f_K}{f_\pi}\frac{F_{B_sK}(M_K^2;0^+)}{F_{B_d\pi}(M_\pi^2;0^+)}
\left(\frac{M_{B_s}^2-M_K^2}{M_{B_d}^2-M_\pi^2}\right),
\end{equation}
we have another interesting probe for $U$-spin-breaking effects. 
Interestingly, the relation $d'e^{i\theta'}=d\,e^{i\theta}$ is not affected 
by $U$-spin-breaking corrections within a modernized version of the 
``Bander--Silverman--Soni mechanism'', making use---among other 
things---of ``factorization'' to estimate the relevant hadronic matrix 
elements \cite{BsKK}. Although this approach appears to be rather simplified 
and may be affected by non-factorizable effects, it strengthens our 
confidence in the $U$-spin relations used for the extraction of $\beta$ 
and $\gamma$ from $B_d\to\pi^+\pi^-$ and $B_s\to K^+K^-$.

The strategy discussed in this section is very promising for 
second-generation $B$-physics experiments at hadron machines, 
where the physics potential of the $B_s$ system can be fully exploited.
At the asymmetric $e^+e^-$ $B$-factories operating at the $\Upsilon(4S)$ 
resonance, BaBar and BELLE, which have already seen the first events, this 
is unfortunately not possible. However, there is also a variant of the 
extraction of $\gamma$, where $B_d\to\pi^\mp K^\pm$ is used instead 
of $B_s\to K^+K^-$ \cite{BsKK}. This approach has the advantage that all 
required time-dependent measurements can in principle be performed at the 
asymmetric $e^+e^-$ machines. On the other hand, it relies---in addition to 
the $SU(3)$ flavour symmetry---on the smallness of certain ``exchange'' 
and ``penguin annihilation'' topologies, which may be enhanced by FSI 
effects. Consequently, its theoretical accuracy cannot compete with the 
``second-generation'' $B_d\to\pi^+\pi^-$, $B_s\to K^+K^-$ approach, 
which is not affected by such problems.

\section{CKM phases and hadronic parameters from angular 
distributions of $B_{d,s}$ decays}

An interesting laboratory to explore CP violation and the
hadronization dynamics of non-leptonic $B$ decays is provided by certain 
quasi-two-body modes $B_q\to X_1\,X_2$ of neutral $B_{d,s}$-mesons, 
where both $X_1$ and $X_2$ carry spin and continue to decay through 
CP-conserving interactions. In a recent paper \cite{RF-ang}, the 
general formalism to extract CKM phases and hadronic parameters from the 
corresponding observables, taking also into account penguin contributions, 
was presented. If we fix the mixing phase $\phi_q$ separately, 
it is possible to determine a CP-violating weak phase $\omega$, which is
usually given by the angles of the unitarity triangle, and
interesting hadronic quantities as a function of a {\it single} 
hadronic parameter. If we determine this parameter, for instance, by 
comparing $B_q\to X_1\,X_2$ with an $SU(3)$-related mode, all remaining 
parameters, including $\omega$, can be extracted. If we are willing to 
make more extensive use of flavour-symmetry arguments, it is possible to 
determine the $B^0_q$--$\overline{B^0_q}$ mixing phase $\phi_q$ as well.

A particularly interesting application of this approach is given by  
$B_d\to J/\psi\,\rho^0$, which can be combined with $B_s\to J/\psi\,\phi$ 
to extract the $B^0_d$--$\overline{B^0_d}$ mixing phase and---if penguin 
effects in the former mode should be sizeable---also the angle $\gamma$ 
of the unitarity triangle. As an interesting by-product, this strategy 
allows us to take into account the penguin effects in the extraction 
of the $B^0_s$--$\overline{B^0_s}$ mixing phase from $B_s\to J/\psi\,\phi$.
Moreover, a discrete ambiguity in the extraction of the CKM angle $\beta$ 
can be resolved, and valuable insights into $SU(3)$-breaking effects can 
be obtained. 

Other applications of the general formalism presented in \cite{RF-ang} 
involve $B_d\to\rho \rho$ and $B_{s,d}\to K^{\ast}\overline{K^\ast}$ 
decays. Since this approach is very general, it can be applied to many 
other decays. Detailed studies are required to explore which channels 
are most promising from an experimental point of view.

\section{Conclusions}

The new strategies discussed above provide an exciting playground 
for second-generation $B$-decay experiments, such as LHCb or BTeV.

\end{document}